\documentclass[runningheads]{llncs}
\usepackage[T1]{fontenc}
\usepackage{graphicx}
\usepackage{url}
\usepackage{color}

\begin{document}
\title{Train While You Fight - Technical Requirements for Advanced Distributed Learning Platforms}
\titlerunning{Train While You Fight - Technical Requirements}
%
\author{Simon Hacks\orcidID{0000-0003-0478-9347}}
\authorrunning{S. Hacks}
%
\institute{Stockholm University, Stockholm, Sweden\\
\email{simon.hacks@dsv.su.se}}
\maketitle              
\begin{abstract}
“Train While You Fight” (TWYF) advocates for continuous learning that occurs during operations, not just before or after. This paper examines the technical requirements that advanced distributed learning (ADL) platforms must meet to support TWYF, and how existing software engineering patterns can fulfill these requirements. Using a Design Science Research approach, we (i) derive challenges from PfPC/NATO documentation and recent practice, (ii) define solution objectives, and (iii) conduct a systematic mapping from challenges to proven patterns. We identify seven technical challenges: interoperability, resilience, multilingual support, data security and privacy, scalability, platform independence, and modularity. We illustrate the patterns with a national use case from the German armed forces.


\end{abstract}
\section{Introduction}\label{sec:introduction}
For decades, military education and training have been guided by the maxim “train like you fight (TLYF).”~\cite{mcgraw2023train} In practice, this meant preparing forces in peacetime through exercises and courseware that emulate anticipated operational conditions as closely as possible, with the expectation that well-rehearsed routines would transfer into real missions~\cite{pfpc2023_info_paper}.

However, recent large-scale conflicts, most prominently Russia’s war against Ukraine, have demonstrated that tactics, techniques, and technologies evolve so rapidly in contact that static, pre-deployment training is no longer sufficient. New weapon systems, counter-measures, and command-and-control (C2) procedures\footnote{C2 describes the structures, processes, and tools by which commanders direct and coordinate forces.} appear and iterate on timescales of weeks or even days, rendering previously prepared materials partially obsolete. This underscores the need for continuous, adaptive learning during operations \cite{pfpc2023_info_paper,presnall2024_twwf}. Reflecting this, the Partnership for Peace Consortium (PfPC) Advanced Distributed Learning (ADL) Working Group Co-Chair and German ADL Director, Lt. Col. Nickolaus, stresses that sustaining effective training through crisis and war is an operational imperative, with the goal of hardening training establishments against intended targeting during war so that personnel can continue learning while fighting.

Train While You Fight (TWYF) addresses these challenges and is operationalized through four complementary strategies~\cite{pfpc2023_info_paper,presnall2024_twwf}: (i) \textit{Lean Principles in Training Design}. Apply pull-driven, just-in-time learning to minimize waste and deliver the right micro-content at the point of need. Streamlined workflows, sequenced prerequisites, and continuous flow reduce cognitive overload and shorten time-to-competency in the theater. (ii) \textit{Mass Customization and Personalized Learning}. Use adaptive learning, modular/reusable learning objects, and configurable scenarios to tailor instruction at scale. Personalized pathways and competency-based progression align training with unit roles, mission profiles, and rapidly changing tactics, techniques, and procedures. (iii) \textit{On-Demand Production of Training Content}. Employ rapid authoring, templated scenario kits, and crowd-sourced subject matter expert (SME) inputs to compress lesson-learned loops. Such as a reach-back workflow, where field users and instructors can submit media and observations to a 24/7 triage cell for rapid author–SME curation and immediate injection into updated learning assets, shrinking the loop from observation to redeployed content. Standardized metadata, localization workflows, and “designed-to-share” licensing accelerate the reuse of content across multiple countries. (iv) \textit{Digital Learning Ecosystems}. Integrate learning management systems (LMS), simulators, performance support, and analytics via standards and federated identity for coalition access. Cloud-hybrid, edge/offline delivery, and Learning Record Stores (LRS) enable resilient, interoperable data flows and secure mobile access under contested conditions~\cite{adl_tla_overview,adl_scorm2004,adl_xapi_103}.

To align with TWYF, classical learning platforms must evolve. Traditional, monolithic LMS deployments and coarse-grained course packages struggle with: (a) interoperability across heterogeneous national systems and standards; (b) resilience to cyber/kinetic disruptions and bandwidth constraints; (c) multilingual localization; and (d) user security and privacy in a federated, cross-border environment. Addressing these constraints requires architectural and design choices that go beyond incremental feature additions~\cite{pfpc2023_info_paper,adl_tla_overview}. 

While technological architectures enable TWYF, it is commanders who operationalize it under pressure. The resilience of training systems depends not only on platforms and networks but on leadership, organizational culture, governance, and institutional readiness to adapt under wartime conditions~\cite{presnall2024_twwf}. Wartime training continuity is bounded by national policies, classification and release rules, organizational silos, and slow administrative cycles. Thus, TWYF demands governance and institutional measures (e.g., catalogs, maturity models, cross-border classification regimes) alongside architecture \cite{mcdc_guidebook_2025,pfpc2023_info_paper,presnall2024_twwf}.

The German Armed Forces are operationalizing TWYF through the \emph{Virtuelle Lernumgebung der Bundeswehr}\footnote{engl.: Virtual Learning Environment of the German Armed Forces} (VLBw), a consolidated, secure digital learning ecosystem intended to replace today’s fragmented solutions with a centralized, modular platform. VLBw is planned to integrate legacy pillars, including Moodle and ILIAS as LMS, a common learning middleware, virtual classrooms (2D/3D), authoring and trainer tools, analytics, and AI-assisted recommendation and search, while enforcing classification-aware access. The aim is “anytime/anywhere/any device” training that tightly couples exercises and operations under a resilient, standards-based architecture.

This paper aims to (i) identify the key technical challenges facing ADL platforms operating under TWYF (Section~\ref{sec:challenges}) and (ii) show how these challenges can be addressed using software engineering patterns (Section~\ref{sec:patterns}). We operationalize this objective through a systematic mapping study (Section~\ref{sec:method}): starting from TWYF-aligned requirements, we elicit challenge statements, then search and screen software engineering pattern corpora to identify patterns that mitigate them, and we discuss how these patterns could be applied using the Bundeswehr’s VLBw as an illustrative use case in the Demonstration (Section~\ref{sec:demonstration}). To avoid duplication and clarify scope, we position this work as the technical enabling layer in a coherent stack: TWYF provides the conceptual paradigm, the MCDC Guidebook supplies the institutional/governance framework, DTRMM offers the measurement and self-assessment mechanism, and this paper contributes the architecture and pattern catalog that operationalize these aims.

\section{Related Work and Background}\label{sec:relatedwork}
\subsection{Train While You Fight}
This section provides a brief overview of ongoing work on TWYF, organized into five broad groups that collectively explain why TWYF is needed and how it can be achieved. 
A first information paper \cite{pfpc2023_info_paper} synthesizes lessons from Ukraine’s wartime use of distributed learning, argues that training is a military target, and recommends resilient-by-design, standards-based ADL with strict handling of personal data, and delivery, shareability, and interoperability across allies and partners under denied, degraded, intermittent, and limited conditions. It frames TWYF as a coalition imperative and lists concrete actions. Moreover, these documents explicitly frame training facilities as deliberate targets.

The next article \cite{presnall2024_twwf} motivates TWYF as a response to rapid operational change, elaborating on “novel strategies”, and highlights the underuse of mature e-learning interoperability standards in multinational sharing. It emphasizes resilient, decentralized learning networks and governance for classification, licensing, and version control, enabling coalition reuse. 

A guidebook \cite{mcdc_guidebook_2025} targets commanders of training establishments, providing a strategic framework, implementation guidance, and tools to sustain training under cyber/kinetic/spectrum pressure. It introduces two enabling artifacts for failover and surge: the Defence Training Resilience Maturity Model (DTRMM) and the Multinational Training Capability Catalog (MTCC), along with templates and checklists to operationalize wartime training resilience. Because DTRMM explicitly recognizes multiple maturity levels, architecture choices can be sequenced as realistic steps; prioritizing minimum viable capabilities first and progressing toward optimization, rather than requiring an immediate leap to full maturity \cite{mcdc_guidebook_2025}.

A concept \cite{pfpc_catalog_2025} proposes a metadata-first, federated catalog for discovering distributed education resources across NATO, allies, and partners. It specifies scope, governance, APIs, controlled vocabularies, resilience measures, and a roadmap toward more than 1,000 entries and strong multilingual coverage.

Complementing the catalog concept, another article \cite{mtcc_concept_2025} defines a two-layer MTCC for physical and digital training capabilities: an unclassified metadata layer for coordination and a classified layer held by nations. It enumerates core fields, use cases, hosting, and governance/integration with existing tools.

\subsection{Research on Advanced Distributed Learning}
ADL represents a modernization of technology-enhanced training and education, moving beyond isolated courseware toward an integrated “learning ecosystem” that spans organizations and platforms \cite{Perisic2023}. Originating as a U.S. Department of Defense initiative, ADL has driven e-learning interoperability and reuse since the early 2000s \cite{adl_scorm2004}. Recent literature highlights that an effective future learning infrastructure will require open interoperability frameworks to connect diverse learning resources and stakeholders globally \cite{Perisic2023}. The ADL vision aligns with these findings by promoting common standards and architectural paradigms that enable lifelong learning and data sharing across traditionally siloed systems \cite{Barr2020}. In practice, ADL’s evolution can be seen in the transition from standalone LMS toward a Total Learning Architecture (TLA) that links multiple systems into a federated, data-driven environment \cite{Smith2021}. This high-level shift—from an LMS-centric model to an ecosystem of interconnected services underpins the technical aspects discussed below.

A key principle of ADL research is modularity. Rather than monolithic platforms, the emphasis is on loosely-coupled services and components that can be independently developed, updated, and recomposed \cite{Smith2021}. For example, the TLA defines a set of functional services with well-defined interfaces, enabling a “plug-and-play” approach to adding or replacing capabilities. Technically, this translates into the adoption of microservice architectures and cloud-native computing in the learning domain \cite{Smith2021}. The benefits are twofold: first, modules (such as content repositories, assessment engines, or recommendation services) can be reused and shared across different training applications; second, the overall system becomes more agile and maintainable, as improvements or new features can be integrated without requiring an overhaul of the entire ecosystem. 

Interoperability is at the core of ADL’s mission, addressed through open standards that ensure different learning tools and content can work together \cite{Barr2020}. The early centerpiece was the Sharable Content Object Reference Model (SCORM), which provided a standardized content packaging format and run-time communication API between courseware and an LMS \cite{adl_scorm2004}. SCORM-enabled reusable content objects and common tracking of learner interactions within traditional web-based courses. However, SCORM had limitations leading to Experience API (xAPI), which allows recording learning experiences in the form of “actor–verb–object” statements, stored in an LRS \cite{adl_xapi_103}. 
Building on xAPI, the cmi5 standard defines how to launch and track content using xAPI within an LMS environment \cite{AICC2016}. It provides a consistent way for content to initiate xAPI tracking upon launch, thereby combining SCORM’s sequencing/packaging concepts with xAPI’s extensible data capture.

Another pillar of interoperability is Learning Tools Interoperability (LTI), an IMS Global (now 1EdTech) standard that allows plug-in integration of external learning tools into platforms \cite{IMS2019}. LTI provides a secure handshake to embed, for example, a third-party simulation or assessment engine into an LMS course, passing context (user identity, course info) seamlessly. Together, these standards form the technical foundation of ADL’s ecosystem \cite{Barr2020}. They enable different systems to exchange data about learners and content in a common language. Current research and development within ADL are also aligning these standards with international efforts to ensure broad adoption \cite{Smith2021}. 


The literature emphasizes the design of training materials for contested and degraded communication environments, meaning that learners at the tactical edge should still be able to access training materials and record progress. On the scalability side, ADL leverages cloud infrastructure and data streaming technologies to handle large volumes of users and data \cite{Smith2021}. By centralizing certain services and federating others, the architecture can scale horizontally as more learning activities and users are added \cite{Barr2020}. A scalable ADL ecosystem also entails performance standards. Early pilot programs demonstrated the ability to index tens of thousands of courses and handle learner data across the force \cite{Barr2020}. These capabilities are crucial not only for efficiency but for enabling advanced techniques like AI-driven personalized training at scale in the future. In sum, resilience and scalability considerations are built into ADL’s design philosophy to guarantee that learning continuity and effectiveness persist from peacetime garrison environments to forward-deployed scenarios.

\section{Research Method}\label{sec:method}
We adopt a Design Science Research (DSR) approach as outlined by Peffers et al.~\cite{peffers2007_dsr}. In this iteration, we complete the phases of (i) problem identification and motivation, (ii) definition of objectives for a solution, and (iii) initial design in the form of a challenge-pattern mapping. We also provide an analytical demonstration that relates the mapping to the VLBw context. Implementation and empirical evaluation are planned for a subsequent DSR cycle.

\textbf{Problem identification and motivation.}
This phase defines the problem space and why it matters~\cite{peffers2007_dsr}. We analyzed the gap between TLYF and TWYF using openly available PfPC and NATO sources, as well as recent practice reports. Concretely, we compiled a corpus of primary documents (PfPC ADL information papers~\cite{pfpc2023_info_paper}, NATO/ADL/TLA handbooks and overviews~\cite{adl_tla_overview}, and TWYF strategies~\cite{presnall2024_twwf}) and performed document analysis to extract statements about constraints, needs, and required capabilities. The result is a consolidated set of seven technical challenges: interoperability, resilience, multilingual support, data security and privacy, scalability, platform independence, and modularity.

\textbf{Define objectives of a solution.}
This phase specifies what a successful artifact should achieve in context~\cite{peffers2007_dsr}. We derived four overarching objectives for a TWYF-aligned platform: (i) enable standards-based interoperability across national systems; (ii) ensure resilience under cyber/kinetic disruption and degraded communications; (iii) provide multilingual delivery and rapid localization; and (iv) enforce user security and privacy in federated, cross-border settings.

\textbf{Design \& development.}
This phase creates the artifact that satisfies the objectives \cite{peffers2007_dsr}. In this iteration, the artifact is a mapping between challenges and possible patterns addressing them: we conduct a review of software engineering pattern catalogs (e.g., POSA \cite{posa1}, GoF \cite{gof1994}, and SEI \cite{sei_security_patterns}) to identify patterns that address each challenge under defense constraints. Inclusion criteria require explicit pattern definition, evidence in distributed or mission-critical contexts, and applicability to NATO/EU standards and governance.

\textbf{Demonstration.}
This phase shows how the artifact addresses the problem in context~\cite{peffers2007_dsr}. In this iteration, we provide a prose demonstration rather than a system prototype: we relate the mapped patterns to VLBw as a national use case, outlining where patterns could be introduced. This demonstration is qualitative and illustrative; no implementation has been performed.

\textbf{Evaluation.}
Evaluation is out of scope for this iteration~\cite{peffers2007_dsr}. We plan a mixed-methods evaluation in the next cycle: (i) technical evaluation via a prototype; (ii) operational evaluation with instructors/trainees in controlled exercises; and (iii) governance/compliance checks for federation, classification, and privacy.

\textbf{Communication.}
We report on the problem framing, objectives, challenge-pattern mapping, and analytical demonstration to the information systems engineering community through this paper and presentation~\cite{peffers2007_dsr}. Future communication will cover the implementation and evaluation results of the prototype. However, some details may remain non-public due to confidentiality.

\section{Challenges of an Advanced Distributed Learning Platform}\label{sec:challenges}

\qquad \textbf{Interoperability}\\
\textit{Challenge.} ADL platforms must integrate heterogeneous national systems (e.g., LMS, LXP, LRS) and exchange content and data across organizational and technical boundaries. Interoperability encompasses syntactic and semantic compatibility, federation of identity and access, and cross-system discovery while respecting classification \cite{adl_tla_overview}. Absent interoperability, content duplication, point-to-point integrations, and fragmented learner records undermine TWYF agility \cite{pfpc2023_info_paper}.\\
\textit{Rationale.} Standards-based interoperability (e.g., SCORM, xAPI, LTI) and cataloging/metadata practices are emphasized~\cite{adl_scorm2004,adl_xapi_103} to enable shareable-by-design content and federated data flows in multinational contexts. Presnall et~al.~\cite{presnall2024_twwf} highlight that allies currently underutilize mature e-learning standards and lack systemic mechanisms for discovery and sharing, creating inefficiencies that TWYF must overcome. The TLA literature~\cite{adl_tla_overview} further motivates the use of interoperable microservice interfaces and common data models to enable components to interoperate without requiring a rip-and-replace approach.

\textbf{Resilience}\\
\textit{Challenge.} TWYF platforms must operate under contested, degraded, and denied conditions, including cyber attacks, kinetic disruptions, intermittent power and bandwidth, and loss of central infrastructure.\footnote{Resilience is not solely technical, but covers multiple layers in an organization, such as human, infrastructure, procedures, and culture. DTRMM formalizes this in six pillars and five maturity levels, so platform choices should be mapped accordingly.}
Resilience requires redundancy, graceful degradation, offline/edge delivery, rapid content replication, and fast recovery while maintaining the integrity and availability \cite{pfpc2023_info_paper}.\\
\textit{Rationale.} Training infrastructure becomes an intended target in war and must be survivable, redundant, and recoverable; PACE-style delivery (Primary/Alternate/ Contingency/Emergency) and cloud–edge hybrids help ensure continuity~\cite{pfpc2023_info_paper}. Presnall et~al.\cite{presnall2024_twwf} argue that resilient, decentralized delivery and rapid incorporation of lessons learned are essential for sustaining learning during operations. ADL/TLA guidance~\cite{adl_tla_overview} similarly recommends architectures that can tolerate component failures and network partitions.

\textbf{Multilingual Support}\\
\textit{Challenge.} NATO coalitions operate in many languages. Therefore, contents must be localized and culturally adapted. This entails end-to-end internationalization of software, scalable localization workflows, metadata in multiple languages, and mechanisms to keep translations aligned \cite{nato_deep_eacademy}.\\
\textit{Rationale.} NATO DEEP~\cite{nato_deep_eacademy} and PfPC~\cite{pfpc2023_info_paper} reports foreground multilingual delivery as a condition for intellectual interoperability and rapid uptake at scale, documenting successful translation campaigns and tooling needs for localization and re-use. Presnall et~al.~\cite{presnall2024_twwf} explicitly call for “designed-to-share” content with clear licensing, classification, and versioning to support multinational localization and economies of scale. Standards bodies and ADL handbooks~\cite{adl_scorm2004,adl_xapi_103} recommend metadata, packaging, and modularity practices that facilitate efficient translation and reassembly of learning objects.

\textbf{Data Security and Privacy}\\
\textit{Challenge.} Learning systems process sensitive personal data (PII), operational contexts, and potentially classified materials; in wartime, compromise of learner identities or activity data can endanger personnel. Platforms must enforce strong authentication and authorization (including federation), protect data in transit and at rest, support auditing, and comply with national/European data protection and data sovereignty requirements across borders \cite{pfpc2023_info_paper}.\\
\textit{Rationale.} PfPC~\cite{pfpc2023_info_paper} stresses the heightened risk of personal data misuse by adversaries and recommends minimizing/omitting PII, enforcing secure-by-design practices, and carefully managing cross-border sharing. Reach-back intensifies security and privacy demands: ingesting media from mobile devices across organizational boundaries requires strong identity and auditable cross-domain handling, while a 24/7 intake hub must absorb bursty uploads from dispersed units and route them through review and publication pipelines without delaying routine delivery. NATO/ADL/TLA guidance~\cite{adl_tla_overview} describes federated identity, role/ attribute-based access control (RBAC), and secure APIs as prerequisites for coalition interoperability that does not erode national control or privacy. Presnall et~al.~\cite{presnall2024_twwf} further emphasize governance (classification, release markings, version control) as intrinsic to shareability and safe reuse.

\textbf{Scalability}\\
\textit{Challenge.} TWYF requires rapidly scaling throughput to train surges of recruits, re-skill units on new systems, and propagate updates within days. Platforms must elastically scale content delivery, assessment, analytics, and storage across regions while sustaining acceptable latency over constrained links \cite{pfpc2023_info_paper}.\\
\textit{Rationale.} Case evidence from Ukraine and NATO exercises points to dramatic fluctuations in training demand and the need for cloud-based elasticity, CDN-style distribution, and edge caches for low-bandwidth theaters~\cite{pfpc2023_info_paper,presnall2024_twwf}. ADL/TLA literature~\cite{adl_tla_overview} motivates the use of horizontally scalable services and decoupled pipelines to absorb spikes without degrading core operations. Presnall et~al.~\cite{presnall2024_twwf} link on-demand content production and mass customization.

\textbf{Platform Independence}\\
\textit{Challenge.} Forces use diverse devices (e.g., desktop, mobile, ruggedized edge), operating systems, and national platforms. To reduce vendor lock-in and maximize reuse, content and services should be portable across environments, and client access should be device- and OS-agnostic with progressive enhancement for denied, degraded, intermittent, and limited conditions \cite{adl_scorm2004,adl_xapi_103}.\\
\textit{Rationale.} Emphasis on standards targets portability across heterogeneous environments and future components, enabling nations to adopt without rip-and-replace \cite{adl_tla_overview}. PfPC~\cite{pfpc2023_info_paper} advocates using open standards and modular interfaces to avoid siloed national solutions and to facilitate federation, including secure mobile access at the point of need. Presnall et~al.~\cite{presnall2024_twwf} underscore cloud–edge hybridity and decentralized delivery as means to reach varied endpoints reliably.

\textbf{Modularity}\\
\textit{Challenge.} To keep pace with change, platforms must support modular design, recomposition of curricula, and selective branching by role, mission, or classification. Modularity reduces coupling between content and delivery channels \cite{adl_scorm2004}.\\
\textit{Rationale.} Contemporary practices (e.g., microlearning, reusable learning objects) consistently argue for modular content to improve adaptability, reuse, and analytics precision \cite{adl_xapi_103}. Presnall et~al.~\cite{presnall2024_twwf} promote mass customization and on-demand production predicated on modular, metadata-rich content and processes. NATO/PfPC recommendations~\cite{pfpc2023_info_paper} further link modularity to efficient translation, classification management, and cross-organizational version control, which are necessary for multinational reuse.

\section{Software Engineering Patterns for an Advanced Distributed Learning Platform}\label{sec:patterns}
In the following, we present the identified patterns and explain how they address the challenges outlined in Section~\ref{sec:challenges}.

\textbf{Interoperability}\\
The \textit{Adapter Pattern}~\cite{gof1994} converts the interface of a component into another interface that clients expect. By using adapters, the platform can wrap legacy LMS, simulation, or repository APIs and present them through a unified interface, allowing otherwise incompatible systems to work together without requiring modifications to their source code. This reduces point-to-point integrations by handling protocol or data-format translation in one place.\\
The \textit{Service Broker Pattern} introduces an intermediary broker to coordinate communication between clients and services to decouple them. In a distributed learning environment, a broker can register available learning services and route requests/responses between them, allowing components to interact via remote calls rather than direct links~\cite{posa1,fernandez2011_esb}. Using a broker/ESB helps content and learner data flow across national systems without bespoke integrations. Moreover, this will enable an effective realization of the reach-back function.\\
\textit{Publish--Subscribe Messaging}~\cite{hohpe2004_eip} decouples producers and consumers by using an message broker. Components publish learning experience data or content updates as messages on a common channel, and other components subscribe to topics of interest. For example, a simulator can emit xAPI statements to a topic that are consumed by an LRS (for recording) and an adaptive tutoring service (for feedback). This asynchronous, many-to-many nature improves scalability and robustness to change and is important for the reach-back functionality.

\textbf{Resilience}\\
\textit{Primary-Backup Replication} maintains a hot-standby instance of a service that can take over when the primary fails, preserving essential functionality in "emergency mode." Primary, backup, and active (state-machine) replication are well-established techniques for fault-tolerant services in distributed systems~\cite{budhiraja1993_primary_backup,schneider1990_state_machine}. For example, a forward-deployed cache/LRS node can temporarily serve content and record events if the central service is unreachable.\\
The \textit{Bulkhead Pattern} isolates resources so that a failure in one component does not cascade to others; the \textit{Circuit Breaker Pattern} prevents repeated failed calls from exhausting resources and enables graceful degradation while a service is down. Both are widely analyzed in the context of microservice resilience~\cite{nygard2018_release_it,mendonca2020_resiliency}.\\
\textit{Client-Side Caching and Synchronization} enables continued learning under denied, degraded, intermittent, and limited conditions by caching content and progress locally and synchronizing with central stores when connectivity is restored, which is well established in distributed systems~\cite{kistler1992_coda,pathan2008_cdn_taxonomy}.

\textbf{Multilingual Support}\\
The \textit{Externalized Resources Pattern} separates all user-facing strings and media from application code into resource bundles, indexed by keys, so that new languages can be added without modifying core logic. Pattern literature on software internationalization documents this approach and related techniques~\cite{mahemoff1999_i18n}. In practice, externalized resources pair well with metadata-driven content pipelines to keep translations aligned with rapidly updated masters.

\textbf{Data Security and Privacy}\\
With \textit{Federated Identity / Single Sign-On (SSO)}, authentication is delegated to trusted Identity Providers, and federation standards enable single sign-on across systems, reducing duplication of PII and aligning security with interoperability needs. Federated identity and access management in multi-organization environments has been extensively studied~\cite{fim4r2013}.\\
\textit{Role-Based Access Control (RBAC)} assigns permissions to roles and maps users to roles/attributes, enforcing least privilege and simplifying audits~\cite{sandhu1996_rbac}.\\
Centralized, tamper-resistant \textit{Audit and Intrusion Detection} hooks at critical points (logins, admin actions, bulk downloads) support incident response and compliance. Design-pattern catalogs for secure systems provide guidance on logging, monitoring, and forensic readiness in distributed environments~\cite{sei_security_patterns}.

\textbf{Scalability}
\textit{Load Balancing} distributes requests across multiple instances to avoid bottlenecks and reduce tail latency; randomized policies such as ``power of two choices'' and modern data-center practices improve robustness~\cite{mitzenmacher2001_two_choices,dean2013_tail_at_scale}.\\
\textit{Caching and Content Distribution} replicate hot data near users to cut latency and conserve bandwidth, which is critical for constrained links~\cite{pathan2008_cdn_taxonomy}.\\
\textit{Sharding and Replication} partition content/learners across shards and use replicas to scale storage. Large systems illustrate design trade-offs for partitioning, replication, and consistency that are directly applicable to LMS and LRS~\cite{decandia2007_dynamo,corbett2012_spanner}.

\textbf{Platform Independence}\\
Encapsulate platform-specific APIs behind a uniform interface (\textit{Wrapper Fa\c{c}ade}) to produce portable code across OS/device differences~\cite{schmidt1999_wrapper_facade}.\\
Expose capabilities via \textit{Microservices} with stable, technology-agnostic APIs so that any device or national platform with a standards-compliant stack can consume them. The microservices body of work discusses decoupling, independent deployment, and the benefits of organizational scalability~\cite{dragoni2017_microservices}.\\
Adopt a \textit{Plugin Architecture} via extension points to add new adapters without changing core logic, separating abstractions from implementations~\cite{posa1,gof1994}.

\textbf{Modularity}\\
Keep a minimal core (\textit{Microkernel}) and add features as plug-ins to enable fine-grained evolution and selective deployment across partners~\cite{posa1}. \\
Decompose into small, independently deployable \textit{Microservices} with API contracts to enable targeted scaling and faster iteration on specific capabilities~\cite{dragoni2017_microservices}.\\
Apply \textit{Separation of Concerns} to keep presentation, domain logic, and integration/data access distinct. Classic modularization principles improve maintainability and allow adaptation to new standards or UIs without ripple effects~\cite{parnas1972_modularity}.

\section{Demonstration}\label{sec:demonstration}
To demonstrate the mapping between challenges and patterns, we use the VLBw as a working example. VLBw is being developed and will be operated on the Bundeswehr private cloud for open and confidential information and assembles a modular ecosystem around a central 2D/3D entry portal, a common learning middleware that integrates functional elements, and a set of core services: LMS, virtual classrooms, a 3D learning environment, authoring and trainer tools, quality-assured document management with workflows, evaluation and exam administration, semantic search with AI-supported recommendations and chatbot, social functions, and a streaming platform. The platform also integrates with systems and training applications, while reusing email, chat, and calendar features from the intranet. Access is differentiated by user groups\footnote{(1) members of the German armed forces; (2) non-members of the German armed forces with access to confidential information; (3) non-members of the German armed forces with no access to confidential information} with integrated registration for linked legacy services.

The entry portal mediates identity and session context to the common learning middleware, which orchestrates traffic between the LMS/virtual classrooms, the 3D environment, and authoring/trainer tools. Content packages, 3D assets, and versioned learning materials are created in the authoring tools and stored under workflow-based QA in the document management service; associated metadata (topic, language, role, classification) is indexed for semantic search. Learner enrollments and course states are managed in the LMS. Session joins and activity events from the 3D environment, which are preferred by technical schools and informed by prior pilots, serve as the primary telemetry stream to update course status and feed evaluation/exam services. 2D classrooms remain available as a fallback and for theory-focused delivery. Social functions and the streaming platform publish and reference training media, which the search service surfaces alongside LMS items. Recommendation and chatbot components utilize usage signals and metadata to suggest next steps. As new native VLBw functions come online, linked legacy systems are incrementally replaced and migrated.

The basic framework is currently implemented and tested for confidential information. A proof of technology at one of the technical schools of the German Armed Forces starts in early 2026, with initial operating capability (IOC) targeted for mid-2026; from IOC onward, the basic framework is rolled out to more schools and training facilities, and in subsequent years, VLBw is expanded to full functionality and introduced at approximately 130 training institutions. In parallel, driven by Russia’s war against Ukraine and the changed security situation, VLBw is expected to more than double its user base, expanding coverage to a larger share of active soldiers, a significantly increased active reserve, and selected blue-light organizations integrated under the national OPLAN\footnote{OPLAN: the national operational plan that defines structures and measures for crisis and war, including the coordinated integration of civilian blue-light organizations with the armed forces.}.

\textbf{Interoperability}\\
A practical first step would be to let the common learning middleware act as a light-weight service broker: register capabilities and broker calls through agreed contracts rather than point-to-point links~\cite{posa1,fernandez2011_esb}. Legacy systems can be wrapped behind Adapter that expose uniform APIs and data schemas, allowing each integration to follow the same template~\cite{gof1994}. For cross-module events, it would be useful to introduce a lightweight publish–subscribe backbone where services emit learning telemetry and content updates on specific topics, and consumers subscribe without tight coupling, by extending the broker with a reach-back intake API and mobile adapters, plus reach-back pub–sub topics~\cite{hohpe2004_eip}.

\textbf{Resilience}\\
To limit blast radius, VLBw services could be deployed in bulkheaded pools (portal, LMS frontends, LRS ingress, search), each with resource limits and isolation boundaries~\cite{nygard2018_release_it}. Calls to external or unstable dependencies (translation engines, external IdPs, remote catalogs) could be guarded by circuit breakers and timeouts with clear fallbacks (cached responses, degraded workflows)~\cite{nygard2018_release_it,mendonca2020_resiliency}. For continuity, forward-deployed cache/LRS nodes are a viable pattern: operate as primary–backup replicas that serve essential content and buffer statements while disconnected, then reconcile on reconnection~\cite{budhiraja1993_primary_backup,schneider1990_state_machine}. Client apps can apply offline-first caching and sync so trainees keep learning and progress is merged later, and the reach-back intake hub is bulkheaded and circuit-broken, with edge/client caching that synchronizes after reconnect.~\cite{kistler1992_coda,pathan2008_cdn_taxonomy}.

\textbf{Multilingual support}\\
Internationalization can be eased by externalizing all UI strings and system messages into resource bundles, allowing new languages to be added without code changes~\cite{mahemoff1999_i18n}. On the content side, a “language-pack” pipeline can keep translations aligned with masters: when a module is updated, the document workflow raises a translation task, preserves IDs/metadata (topic, role, classification, license), and blocks publication until required locales are ready.

\textbf{Data security and privacy}\\
For cross-organization access, the portal could consume assertions from partner IdPs and enable SSO, which reduces PII replication and simplifies account lifecycle~\cite{fim4r2013}. Authorization can rely on RBAC with attributes to express coalition policy in simple, auditable rules~\cite{sandhu1996_rbac}. All modules should emit tamper-resistant audit logs to a central monitor with alert rules for sensitive actions~\cite{sei_security_patterns}. Classification and release markings should be propagated as data tags across services and enforced at service boundaries, with reach-back access gated by federated SSO and RBAC, on-device redaction, and audited cross-domain guards.

\textbf{Scalability}\\
Frontends (portal, LMS) will benefit from load balancers; simple “power-of-two-choices” selection helps avoid hot spots during surges~\cite{mitzenmacher2001_two_choices,dean2013_tail_at_scale}. Large media and popular content should be pushed to edge caches at school sites to reduce latency and backhaul~\cite{pathan2008_cdn_taxonomy}. LRS and content stores can be sharded by school/region/tenant with replicated readers to scale analytics and reporting horizontally~\cite{decandia2007_dynamo,corbett2012_spanner}. Where workloads burst (e.g., certificate generation, statement post-processing), queue-based workers can smooth peaks~\cite{hohpe2004_eip}.

\textbf{Platform independence}\\
On the client side, a \emph{Wrapper Fa\c{c}ade} can hide device specifics so core logic remains portable across desktop, mobile, and rugged laptops~\cite{schmidt1999_wrapper_facade}. On the server side, capabilities should be exposed as stable microservice APIs so national tools or future clients integrate through plugin-style adapters without changes to core services, while mobile capture uses a wrapper fa\c{c}ade and plugin adapters, and reach-back itself is exposed as stable APIs.~\cite{dragoni2017_microservices,posa1,gof1994}.

\textbf{Modularity}\\
The portal plus middleware can evolve toward a microkernel, keeping identity, navigation, policy, and eventing in a minimal core, and plugging in LMS, exams, forums, a 3D world, search, and recommendation through clear extension points~\cite{posa1}. Services should be kept small and independently deployable (microservices) with strict boundaries and \emph{separation of concerns} to localize change and allow stepwise replacement of legacy components~\cite{dragoni2017_microservices,parnas1972_modularity}.

To make the architecture actionable for commanders, the platform needs to provide “decision hooks” that translate technical signals into choices during disruption, relocation, or degraded operations. A minimum viable training standard dashboard could provide indicators such as the share of critical modules reachable, LRS ingestion lag, and tail latency to distinguish acceptable degradation from mission-impairing loss. Relocation should be supported by federated-identity readiness checks and a capability catalog to identify alternative facilities with the right classification and bandwidth, while the service broker rebinds endpoints and pre-positions content to edge nodes.

\section{Discussion}\label{sec:discussion}
This study presents a design-oriented mapping of patterns to challenges, but its interpretation is shaped by several boundaries that affect how the results should be understood. Our problem framing relies on openly available PfPC/NATO documents and public technical standards; classified operational details and nation-specific constraints were outside our reach. That choice keeps the work reproducible but may miss requirements that only surface in protected settings. The coding of documents and consolidation of statements reflect our interpretation of terms such as resilience, platform independence, or multilingual support, which raises the risk of construct bias even though we applied explicit inclusion criteria. We also assume broad availability and consistent interpretation of standards. In practice, divergent national profiles and policy restrictions can limit interoperability and sharing, and these frictions are only partly visible in open sources. Finally, our focus is technical; governance, acquisition, sustainment, and training-of-trainers are acknowledged but not evaluated, even though they strongly influence outcomes.

The analysis emphasizes established distributed systems, integration, and security patterns. This keeps the mapping grounded, yet it is not exhaustive: other families (e.g., formal assurance methods, domain-specific e-learning workflows, edge AI) are only touched indirectly. Our reasoning is architectural and qualitative; we do not present performance envelopes, failure-mode analyses, or cost-of-change models. As a result, causal claims of the form “pattern mitigates challenge” remain argued from literature and engineering judgment rather than from controlled comparisons. Real-world circumstances, such as network topology, spectrum denial, legacy lock-in, and data classification policy, can significantly influence outcomes and may require different compositions than those suggested here. The demonstration is analytical and tied to the intended VLBw architecture; without a prototype, field trials, or user studies, we cannot make claims about effectiveness, effort, or risk reduction in practice.

Generalization and reliability also merit caution. VLBw provides a concrete and valuable context, but architectural baselines, legal frameworks, and operational conditions vary across nations and coalitions, which limits external validity. We make no statistical claims and report no effect sizes; the absence of quantitative benchmarks weakens the validity of our conclusions.

\section{Conclusions}\label{sec:conclusions}
This paper aims to understand what advanced distributed learning platforms must provide to support 'Train While You Fight' and how proven software engineering patterns can contribute to this goal. From PfPC/NATO sources and recent practice, we distilled seven technical challenges and mapped each to concrete patterns. We then suggested how these patterns could be introduced in VLBw, outlining information flows and a staged rollout.

The main insight is that TWYF does not require entirely new technology; it needs disciplined use of known patterns composed to meet coalition constraints. A brokered, event-driven core with a strong identity and data tagging creates room for rapid change without requiring a rip-and-replace approach. Edge-aware delivery and synchronization enable continuous learning even when network links are weak. Clear boundaries reduce coupling, so nations can adopt pieces at their own pace while still participating in a federated ecosystem. Treating training infrastructure as an intentional target clarifies design priorities as first-order requirements. Composed per our pattern blueprint, these mechanisms expose commander-visible decision hooks, aligning technical feasibility with governance in disruption scenarios.

This work has limitations: it is based on public documents, employs qualitative reasoning, and presents an analytical demonstration. As a result, effects and trade-offs are debated, rather than measured. Future work will focus on prototyping the suggested solution, defining and collecting quantitative metrics, and conducting user-facing exercises to assess operational fit. We also plan cross-national interoperability trials, red-team assessments for security and data governance, and an extended pattern catalog covering AI-assisted content pipelines, observability, policy-as-code, and zero-trust variants. The goal is to establish a reference implementation and evidence base that transforms the proposed blueprint into a validated path for resilient multinational learning under TWYF.

\begin{credits}
\subsubsection{\ackname} I would like to thank Dr. Aaron Presnall and Lt. Col. Michael Nickolaus for their valuable input on this work.

\subsubsection{\discintname}
The author serves as a reserve soldier in the unit overseen by Lt. Col. Michael Nickolaus, which is responsible for the VLBw program. This study was conducted entirely outside that service. The Bundeswehr provided no study-specific funding and had no role in the study design, data collection, analysis, writing, or the decision to publish. Only publicly available and unclassified sources were used.
\end{credits}
%
%
%
\bibliographystyle{splncs04}
\bibliography{mybibliography}
\end{document}